\documentclass[aip,rsi,reprint]{revtex4-1}

\draft 

\usepackage{siunitx}
\usepackage{graphicx}
\usepackage{amsmath}
\usepackage[dvipsnames]{xcolor}
\usepackage{xurl}
\begin{document}

\title{An Ultra-High Vacuum Scanning Tunneling Microscope with Pulse Tube and Joule-Thomson cooling operating at sub-pm $\boldsymbol{z}$-noise 
    }

\author{Marcus Eßer}
\affiliation{2nd Institute of Physics B, RWTH Aachen University, 52074 Aachen, Germany}
\author{Marco Pratzer}
\affiliation{2nd Institute of Physics B, RWTH Aachen University, 52074 Aachen, Germany}
\author{Marc Frömming}
\affiliation{2nd Institute of Physics B, RWTH Aachen University, 52074 Aachen, Germany}
\author{Jonas Duffhauß}
\affiliation{2nd Institute of Physics B, RWTH Aachen University, 52074 Aachen, Germany}
\author{Priyamvada Bhaskar}
\affiliation{2nd Institute of Physics B, RWTH Aachen University, 52074 Aachen, Germany}
\author{Michael. A. Krzyzowski}
\homepage[]{https://www.cryovac.de/}
\affiliation{CryoVac GmbH \& Co KG, 53842 Troisdorf, Germany}
\author{Markus Morgenstern}
\email[]{Corresponding author: mmorgens@physik.rwth-aachen.de}
\homepage[]{https://www.institut2b.physik.rwth-aachen.de}
\affiliation{2nd Institute of Physics B, RWTH Aachen University, 52074 Aachen, Germany}

\date{\today}

\begin{abstract}
We describe a compact ultra-high vacuum (UHV) scanning tunneling microscope (STM) system that does not need any external supply of cooling liquids. It achieves temperatures down to \SI{1.5}{\kelvin} and a $z$-noise down to \SI{300}{\femto\meter_{RMS}} for the frequency range of \SI{0.1}{Hz} - \SI{5}{\kilo\Hz} (feedback loop off). It employs a pulse tube cryocooler (PTC) and a Joule-Thomson (JT) stage inducing only small temperature oscillations at the STM with amplitude below \SI{1}{\milli\kelvin}. The challenge to combine an effective vibrational decoupling from the PTC with sufficient thermal conduction is tackled by a multipartite approach. We realize a minimal stiffness of the UHV bellows that connect the PTC and the STM chamber. Fine Copper wires mechanically decouple the PTC stages from cooling plates that carry the thermal shields, the JT stage and the STM. Soft springs decouple the STM from the JT stage. Finally, the STM body has an optimized conical shape and is made of the light and stiff material Shapal\texttrademark\ Hi MSoft such that a strong reduction of low frequency vibrations results for the tunnel junction. The voltage noise in the tunnel junction is 
\SI{120}{\micro\volt} and an RF antenna close to the tunnel junction provides radio frequency excitations up to \SI{40}{\giga\Hz} with amplitudes up to \SI{10}{\milli\volt}.
\end{abstract}


\maketitle 

\section{Introduction}\label{sec1}

	 Scanning tunneling microscopy (STM) and spectroscopy (STS) are central tools in nanoscience. They are used, e.\,g., to obtain a microscopic understanding of electronic phenomena down to the atomic scale.\cite{Eigler1990,Crommie1993} Cryogenic temperatures improve the required energy resolution, that is better than 1\,meV below 4\,K,\cite{Ast2016,Tersoff1985} and the instrumental stability
     such as the thermal drift being typically below \SI{1}{\pico\meter\per\min}. \cite{Liebmann2017} A low temperature is also required to stabilize self-made atomic structures against diffusion,\cite{Eigler1990,Crommie1993} and to get access to low-temperature phenomena \cite{Elrod1984,Morgenstern2004} such as superconductivity \cite{Fischer2007, Hanaguri2010,Yazdani1997} including Majorana modes,\cite{Perge2014} the quantum Hall effect,\cite{Hashimoto2008,coissard2022,Liu2022},  Luttinger liquids,\cite{Blumenstein2011,Jia2022,Jolie2019} or the Kondo effect \cite{Madhavan1998,Wahl2007,Li1998}. 
  
  A major threat are the continuously rising helium prices.\cite{USGeologicalSurvey2001,USGeologicalSurvey2023} While the worldwide helium demand continuously increases,\cite{Maura2023} e.g. predicted to grow by $\sim$\SI{50}{\percent} until 2040,\cite{Provornaya2022} frequent reports describe a negatively counteracting global helium shortage.\cite{Rosen2023,Appel2022,Böck2022,Seidler2019,Hopkins2023} Since the helium gets partially lost ($\sim$\SI{5}{\percent}), if  provided by a combination of external liquefier and helium recovery system, closed-cycle cooling techniques became increasingly popular as they prevent such losses. 
  
As an additional advantage, the closed-cycle operation eliminates the helium refilling, that usually interrupts the measurements and requires educated personnel. Closed-cycle operation also prevents small changes in temperature that can otherwise be caused by a changing filling level of helium in the cryostat. For STM, the refilling is, moreover, accompanied by a risk of a microtip change and typically induces undesired thermal drift of the tip with respect to the sample.  Hence, closed-cycle operation without refilling enables more long-term, rather stable experiments virtually without helium losses.

However, for STM, the closed-cycle cooling is challenging since the typically used pulse tube coolers (PTCs) operate with pressure pulses at roughly 1\,Hz that must be vibrationally decoupled from the STM. 
Consequently, STM closed-cycle operation lags significantly behind other experimental methods.
The first home-built STM system cooled by a commercial PTC worked with an additional dilution refrigerator, but without ultra-high vacuum (UHV), and achieved atomic resolution on highly oriented pyrolytic graphite (HOPG) at 15 mK, but did not specify the $z$-noise for STM.\cite{Haan2014} Later,  commercial systems consisting of a PTC and a variable temperature insert have been employed to immerse the STM in liquid helium. This led to a $z$-resolution in the low pm range at 1.6\,K.\cite{Meng2019,Geng2023} Within UHV, the first systems employed only the PTC for cooling achieving down to 4.6\,K with the help of an exchange gas that thermally couples the PTC to the STM.\cite{Hackley2014, Zhang2016,Coe2024, Kasai2022} Such systems are also commercially available by Unisoku \cite{Kasai2022}, Scienta Omicron and RHK Technology \cite{Chaudhary2021}. Only recently, additional cooling stages have been added for achieving lower temperatures in UHV.\cite{Huang2022,Ma2023} One strategy was to externally produce the liquid helium, but tightly connect the closed-cycle flow of He to the STM chamber, where an additional Joule-Thomsen (JT) stage was used to achieve 3\,K at the STM head.\cite{Ma2023} Another, principally more compact design integrated the PTC into the STM chamber for filling a 1K pot by the produced liquid He that during subsequent pumping enabled 1.4\,K at the STM.\cite{Huang2022} In the former case, we have estimated the $z$-noise obtained without feedback from the displayed current noise to $\sim$\SI{2}{\pico\meter_{RMS}} at 1\,kHz bandwidth. In the latter case, only data with feedback are presented exhibiting sub-pm noise, which, however, cannot be compared to other systems since depending in detail on the unknown feedback parameters that counteract the vibrations in constant-current mode.

Here, we present a UHV-STM with integrated PTC and a connected JT stage. It reliably exhibits sub-pm $z$-noise at 1.5\,K without feedback, partially going down to 300 fm$_{\rm RMS}$. It operates with a PTC from TransMit (PTD 411 on a DN160CF flange) that has the advantage to be bakeable up to $130^\circ$\,C, tuneable in terms of operation frequency (1.1\,Hz - 1.6\,Hz), and connected to the rotary valve by one flexible tube only. For efficient vibrational decoupling, the two PTC stages are each thermally coupled to a separate cooling plate by bundles of 50\,µm thin Cu wires. The cooling plates are directly connected to thermal shields and the JT stage that eventually holds the STM via three soft springs. The cooling plates are carried by a honeycomb-type construction with high stiffness and low thermal conductivity that is directly screwed to the UHV-chamber surrounding the STM. To decouple this chamber vibrationally from the rigidly mounted PTC, we employ UHV bellows with minimal stiffness achieved by counteracting springs.\cite{Li2020, Kirk1978}  Thus, the UHV system supported by Minus K\textsuperscript{\textregistered} vibrational isolators moves in all directions with a resonance frequency that is below the operation frequency of the PTC.
Finally, we designed the STM body to obtain very low transmission of low frequency vibrations. The transmission to the tunnel junction is measured as $10^{-5}$ at 10\,Hz and extrapolated to $10^{-7}$ at 1\,Hz. 

To demonstrate the STM performance, we present atomic resolution and standing electron waves on Au(111), superconducting gaps of Pb for electronic temperature calibration (1.7\,K), Pb-Pb Josephson currents to determine the remaining voltage noise (120\,$\mu$V) and microwave assisted Cooper pair tunneling to evidence high frequency operation of the tunnel junction up to 40\,GHz. 
 
\section{The Setup}\label{sec2}

\subsection{Concept}\label{subsec2_1}

\begin{figure*}[hbt] 
\includegraphics{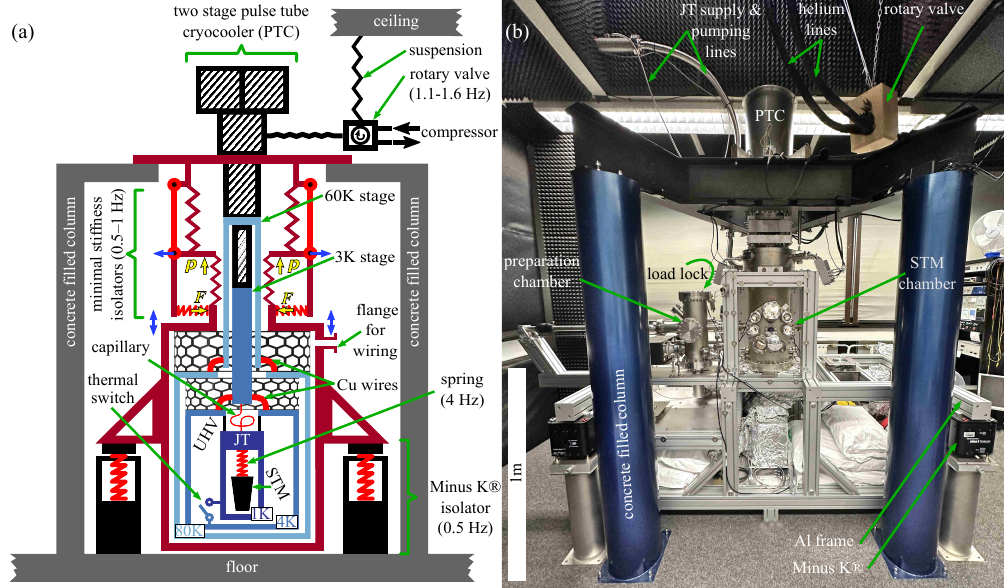}
\caption{\label{fig:UHV System} {\bf Concept of the system.} (a) Schematic cross section highlighting the  vibrational isolation. The UHV chamber with the STM is connected to the PTC only via Cu wires and via the two bellows (red zigzag lines) that are tuned to low resonance frequencies (0.5-1\,Hz) by compensating springs (Fig.~\ref{fig:minimal stiffness}). Consequently, the chamber, that is supported by three Minus K\textsuperscript{\textregistered} isolators,  moves vertically and horizontally at low eigenfrequency (blue arrows). (b) Photograph of the system.}
\end{figure*}

A schematic cross section of the system is depicted in Fig.~\ref{fig:UHV System}a. Figure~\ref{fig:UHV System}b shows a corresponding photograph. The PTC with a cooling power of 1\,W at 4\,K and 20\,W at 60\,K operates with helium pressure pulses at tunable frequency of 1.1-1.6\,Hz and is, thus, a major source of vibrations. Hence, the PTC is mounted on a rigid support frame (double-T beam steel construction) that itself rests on three concrete filled columns standing on the ground floor of the building. The helium pulses are induced via a rotary valve connected to the PTC with only one flexible tube of stainless steel. The valve is hanging from the ceiling and linked to the helium compressor, that is located within a separate room, by two helium lines. To reduce acoustic noise, these lines are wrapped in  butyl rubber and neoprene fabric and the rotary valve is housed in a wooden box with the inner walls covered by butyl rubber and acoustic foam (Fig.~\ref{fig:UHV System}b). For the same reason, the PTC head is enclosed by a stainless steel tube filled with stone wool. 

The crucial task is to decouple these vibrating parts from the UHV chamber that houses the STM, while still enabling sufficient cooling power. Thus, the thermal connection is realized via thin Cu wires (diameter: 0.05\,mm) towards cooling plates that are suspended from the top of the STM chamber by a honeycomb structure (black honeycomb structure in Fig.~\ref{fig:UHV System}a). This structure consists of two consecutive stainless steel cylinders with thin walls. The cylindrical structure is optimized for high stiffness and low thermal conductivity (Fig.~\ref{fig:grid structure}). The cooling plates carry the radiation shields for each temperature and the JT stage that holds the STM via three damped springs. The helium supply of the JT stage is realized by a soft capillary tube (Fig.~\ref{fig:UHV System}a) that is precooled by the two PTC stages (Fig.~\ref{fig:Cryostat performance}c). The helium pumping line from the JT stage is mechanically not connected to the PTC. It is guided through a flange of the suspended UHV chamber to a flexible bellow connecting to a turbomolecular pump. For thermal decoupling, the JT stage is suspended from the 4\,K cooling plate by only three small cross section stainless steel tubes (black rectangles in Fig.~\ref{fig:UHV System}a).

The whole UHV chamber is then mechanically  decoupled from the PTC by reducing the stiffness of two interconnecting UHV bellows with the help of counteracting external spring forces that can be tuned in strength via screws (Fig.~\ref{fig:UHV System}a, \ref{fig:minimal stiffness}a). 
Generally, these spring forces counteract all restoring forces acting on the complete UHV system, including an Al frame,  that floats on three Minus K\textsuperscript{\textregistered} isolators (Fig.~\ref{fig:UHV System}b). The counteracting forces are tuned while monitoring the resonance frequencies of the floating UHV system in the three different, independent directions. This favorably results in frequencies below the operation frequency of the PTC (Fig.~\ref{fig:UHV System}a, \ref{fig:minimal stiffness}d), realizing a minimal stiffness support that is often dubbed negative stiffness concept.\cite{Platus1992, Ibrahim2008, Li2020} Such concept has also been used for the vibrational preisolation in gravitational wave detectors \cite{Blair1993, DeSalvo2007} or during the development of the James Webb Space Telescope to create a flight-like environment by suspending the world largest vacuum chamber \cite{Feinberg2014}.
Finally, the electric wiring for the STM is guided directly to the suspended UHV chamber (Fig.~\ref{fig:UHV System}a) to avoid vibrationally induced electrical noise.\cite{Kalra2016} 

 Besides the STM chamber, the UHV system consists of a preparation chamber, a load lock and a combined prepumping line with a turbomolecular pump and a mass spectrometer (Fig.~\ref{fig:UHV System}b). They are all located on the same Al frame as the STM chamber and, hence, belong to the part that is decoupled from the PTC by the two soft bellows and the three Minus K\textsuperscript{\textregistered} damping stages. The chambers can be pumped and baked-out separately and subsequently maintain their pressure by individual ion getter pumps. In the STM chamber, the geometry of viewports, mirrors and STM head are designed for a 45$^\circ$ optical view onto the sample inside the STM. This enables a tip approach on exfoliated flakes under optical control with $\sim$\SI{10}{\micro\meter}  resolution.\cite{Geringer2009,Johnsen2023}

The whole STM system is surrounded by sound-absorbent walls and separated from the STM control unit (and occasionally the operator) by a sound-insulating wall. The helium compressor (Sumitomo\textsuperscript{\textregistered} F-70H) is in an adjacent room, where it is placed on a rubber mat inside a stone wool damped fibreboard box. The pumps and the helium bottle, needed for the operation of the JT stage, are in a separate room. They are connected to the UHV system via flexible JT supply and pumping lines (Fig.~\ref{fig:UHV System}b).

 	\subsection{Cryogenic Performance}\label{subsec2_2}
\begin{figure*} 
\includegraphics{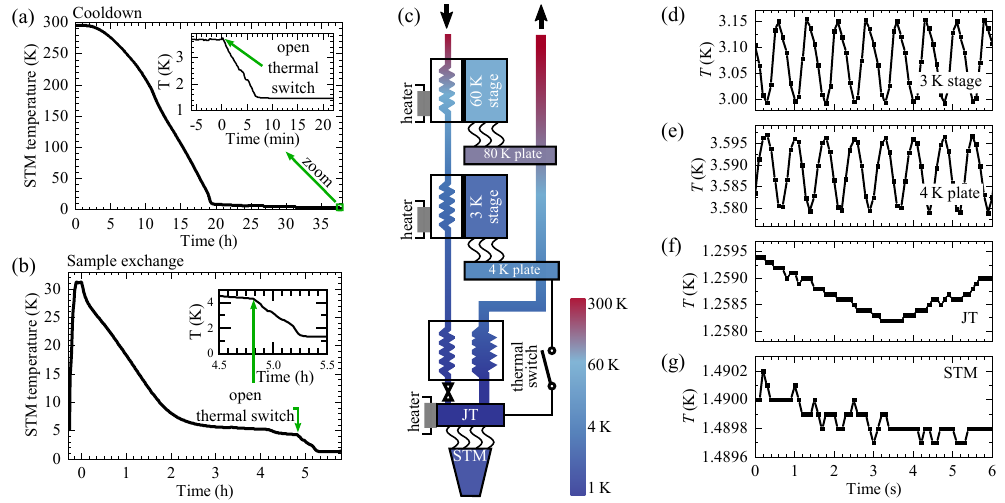}
\caption{\label{fig:Cryostat performance} {\bf Cryostat performance.} (a) Temperature at the STM during an initial cooldown from 300\,K, inset: zoom with marked opening time of the mechanical switch between 4\,K shield and JT stage (Fig.~\ref{fig:UHV System}a). (b) Temperature development of the STM during cooling after exchanging the sample, the opening time of the thermal switch is marked, inset:  zoom into the main image. (c) Schematic of the cryostat colored with the temperatures during operation, the 80\,K plate and 4\,K plate are connected  to the PTC stages by fine copper wires. (d)--(g) Temperature oscillations during PTC operation measured at various stages as marked. }
\end{figure*}

 The two-stage PTC provides 80\,K and  4\,K at the two cooling plates (Fig.~\ref{fig:Cryostat performance}c) as measured by a silicon diode and a resistance thermometer, respectively (Lake Shore Cryotronics\textsuperscript{\textregistered}, Inc. DT-670-A1-CO and CX-1030, respectively).
  Heat exchangers at the PTC stages enable the cooling and liquefaction of helium gas for the JT stage. This helium is guided by a soft capillary that is additionally shielded against radiation by coaxial tubes connected to the PTC stages only and not to the cooling plates. The heat exchangers and the JT stage are equipped with heaters for simulating heat loads and, hence, measuring the cooling power or for operating the STM at elevated temperature.
 The JT stage can be thermally connected to the 4\,K plate by a mechanical switch (thermal switch, Fig.~\ref{fig:UHV System}a) for precooling.
After decoupling, it achieves 1.2\,K as measured by a Cernox\textsuperscript{\textregistered} sensor (Lake Shore  Cryotronics\textsuperscript{\textregistered}, Inc., CX-1030-HT-0.3M). An impedance at the JT stage realizes the pressure drop that eventually provides the cooling power. The pumped helium after the JT stage is guided through an additional counterflow heat exchanger for precooling the incoming liquid (Fig.~\ref{fig:Cryostat performance}c).\cite{Zhang2011} The two cooling plates and the JT stage each carry a radiation shield  (Fig.~\ref{fig:UHV System}a) with a shutter that can be opened for exchange of samples and tips in the STM.

 Starting at room temperature, the STM is cooled to base temperature within roughly 40\,h (Fig.~\ref{fig:Cryostat performance}a). The base temperature at the STM is 1.5\,K as measured by a Cernox\textsuperscript{\textregistered} sensor  (Lake Shore Cryotronics\textsuperscript{\textregistered}, Inc.  CX-1030-HT-0.3M). 
For precooling, the JT stage is connected to the 4\,K plate by the thermal switch (Fig.~\ref{fig:UHV System}a).
The switch remains kept closed until the JT stage reaches $\sim$\SI{4}{\K}. After opening the switch, it takes additional $5$\,min to reach $T=1.5$\,K at the STM (inset, Fig.~\ref{fig:Cryostat performance}a). 
 During sample or tip exchange, the subsequent lateral positioning of the sample and the coarse approach of the tip, the STM typically heats up to $\sim$\SI{30}{\K}. In that case, the base temperature is recovered within $\sim$\SI{5}{}\,hours (Fig.~\ref{fig:Cryostat performance}b). 
 
 One adverse property of cryocoolers are temperature oscillations caused by the pulsed gas and, hence, pulsed cooling power.\cite{Pizzo2022} In our system, they amount to  \SI{150}{\milli\kelvin_{pp}} directly at the 2$^{\rm nd}$  stage of the PTC (3\,K stage, Fig.~\ref{fig:Cryostat performance}d), but are reduced to 15\,mK$_{\rm pp}$ at the adjacent cooling plate (4\,K plate, Fig.~\ref{fig:Cryostat performance}e) and to less than \SI{1}{\milli\kelvin_{pp}} at the JT stage and the STM (Fig.~\ref{fig:Cryostat performance}f-g). 

 \subsection{Vibrational Decoupling of the UHV Chambers}

 \begin{figure} 
\includegraphics{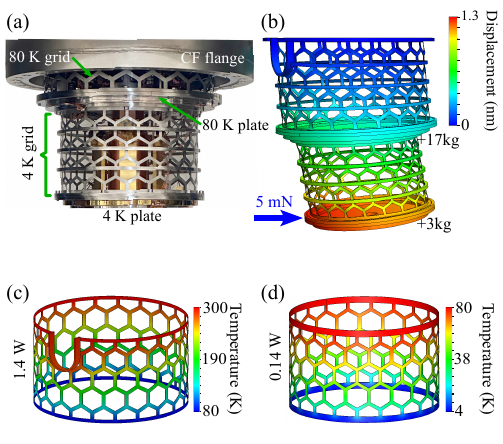}
\caption{\label{fig:grid structure}  {\bf Support structure for the cooling plates.} (a) Photograph of the lower part of the honeycomb-type support. The upper part is partly hidden by the flange. (b) Simulated static displacement of the honeycomb structure for a force of 5\,mN as marked. The force is estimated from the measured vibrational amplitude of the PTC stage and the stiffness of the interconnecting  Cu wires. Weights are added homogeneously to the plates as marked to approximate the influence of the thermal shields and other weights. The static displacement emulates the response well below the simulated resonance frequency of the bending mode of the structure (60\,Hz). (c)-(d) Simulated temperature distributions with fixed temperatures at the top and the bottom of each cylinder. The resulting heat flow across the cylinder is marked on the left. }
\end{figure}

The connection of the cooling plates to the PTC stages by only fine Cu wires requires a rigid support of the plates. The plates carry the thermal shields, the JT stage and the STM amounting to 17\,kg and 3\,kg at the 80\,K and the 4\,K plate, respectively (Fig.~\ref{fig:grid structure}b). 
The support structure is depicted in Fig.~\ref{fig:grid structure}. It is made rigid enough to avoid vibrational excitations by the forces that act on the plates via the Cu wires.
Moreover, its thermal conduction is made low enough such that the PTC can  cool the plates sufficiently. We manufactured a honeycomb-type grid structure from a stainless steel sheet (thickness: 0.5\,mm) that is subsequently bent into a cylinder and spot welded by a laser (Fig.~\ref{fig:grid structure}a). Stainless steel enables low  thermal conduction and has favorable machining properties. To increase the stability against bending, stainless steel rings 
surround the cylinder in regular vertical distance (Fig.~\ref{fig:grid structure}a-b). They are spot welded to each of the vertical beams of the honeycomb structure. 
The first cylinder is fixed at the top flange of the UHV chamber and carries the 80\,K plate. The second cylinder is fixed to the 80\,K plate and carries the 4\,K plate (Fig.~\ref{fig:grid structure}b). 

For the optimization of the design, we employed finite element simulations using SOLIDWORKS\textsuperscript{\textregistered}. As input, we measured the vibrational amplitudes of the 3\,K PTC stage and extrapolated it to the point, where the Cu wires are fixed. At the operation frequency of the PTC, the amplitudes amount to $\sim$\SI{30}{\micro\meter} in vertical and $\sim$\SI{10}{\micro\meter} in horizontal direction. 
Using the stiffness of the Cu wires, we estimated the resulting force amplitude acting on the cooling plate to be $\sim$\SI{5}{\milli\newton} in the bending direction (Fig.~\ref{fig:grid structure}b) and $\sim$\SI{10}{\milli\newton} in the vertical direction (not shown). The simulation then yields vibrational amplitudes of the cooling plate of $\sim$\SI{1}{\nano\meter} in both directions as shown for the horizontal one in Fig.~\ref{fig:grid structure}b. This indicates a more than sufficient stiffness of the construction.

\begin{figure*} 
\includegraphics{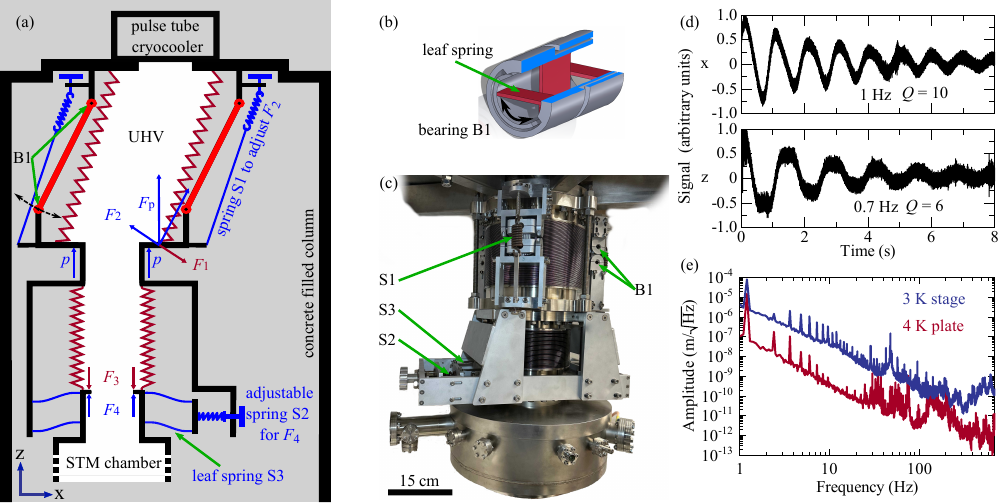}
\caption{\label{minimal stiffness setup}  {\bf Minimal stiffness setup for vibrational decoupling.} (a) Schematic of the two bellows (red zigzag lines) and the compensating springs S1-S3 with indicated counteracting forces (blue). The lower bellow is pushing downwards due to an upwards compression with respect to its equilibrium ($F_3$) which is counteracted by the preloaded  leafsprings ($F_4$). The counteraction works similarly, but in opposite direction, for an extended bellow. The horizontal spring S2 sets the preload until one gets a low vertical resonance frequency, i.e. minimal stiffness in vertical direction. In the upper part, the stiffness of the bellow drives the laterally displaced STM chamber back into its equilibrium position ($F_1$). This is counteracted by the external pressure $p$ pushing onto the lower flange  ($F_p$) and the adjustable spring force $F_2$ that can be tuned until the vertical (or pendulum like) resonance frequency is low. (b) Drawing of the frictionless bearings that are used in the upper part of a at the joints of the red rods. (c) Photograph of the assembled negative stiffness setup with some marked bearings B1 and springs S1-S3. (d) Resulting movements of the UHV system floating on the Minus K\textsuperscript{\textregistered} isolators (Fig.~\ref{fig:UHV System}) after optimization of the compensating spring forces and initial manual displacement in horizontal $x$  (top) and vertical $z$ direction (bottom). (e) Frequency spectrum of vibrational amplitude in vertical direction recorded at the 3\,K PTC stage and the 4\,K cooling plate without the JT stage/STM. 
\label{fig:minimal stiffness}}
\end{figure*}

The heat transfer was simulated by finite element simulations in steady state (Fig.~\ref{fig:grid structure}c-d). We find a heat load of 1.4\,W from 300\,K towards the 80\,K plate. This is only 10\,\% of the cooling power of the corresponding PTC stage at the measured temperature of this stage (62\,K) during operation with the whole system installed. A much stronger heat load originates from the thermal radiation absorbed by the corresponding radiation shield. We estimate this radiation load to 9\,W using an absorption coefficient $\varepsilon_{\rm Al} =0.03$ for the polished Al of the shield \cite{Ekin2006} and an emission coefficient $\varepsilon_{\rm steel}=0.15$ for the surrounding stainless steel UHV chamber that is at $300$\,K \cite{white2002}.   
At the 4\,K plate, we get a heat load of 0.14\,W from the 80\,K plate. This amounts to $\sim$\SI{70}{\percent}
of the cooling power of the corresponding PTC stage at the measured 2.8\,K. 
Using the 0.14\,W as the only heat load, the PTC calibration predicts a PTC stage temperature of 2.6\,K, well below the required 4\,K for liquifying the helium of the JT stage.
Hence, the cylindrical honeycomb structure indeed provides low enough thermal conduction at a sufficient stiffness.

In order to profit from the soft mechanical coupling between PTC stages and cooling plates via the fine Cu wires, we must avoid mechanical shortcuts along other paths. The most stiff connection is provided by the UHV bellows between the PTC head and the UHV chamber containing the STM (Fig.~\ref{fig:minimal stiffness}a). The bellows have been chosen as soft as possible, but still provide relatively strong restoring forces, with effective spring constants of 2.5\,kN/m and 6\,kN/m in the vertical and the horizontal direction, respectively. Since the whole UHV system is floating on the Minus K\textsuperscript{\textregistered} isolators, the bellows cause the dominating restoring forces for the whole UHV system including its supporting Al frame (Fig.~\ref{fig:UHV System}).    
Other restoring forces originate from the external JT pumping line (Fig.~\ref{fig:UHV System}b), the internal He capillary towards the JT stage and the Cu wires from the PTC stages to the cooling plates (Fig.~\ref{fig:UHV System}a). But they contribute to a lesser extent as cross checked experimentally. All these restoring forces are compensated simultaneously by two sets of counteracting springs (Fig.~\ref{minimal stiffness setup}a) located at the two bellows. The springs can be tuned to realize low stiffness in vertical and the two horizontal directions, respectively. 

 \begin{figure*}[hbt] 
\includegraphics{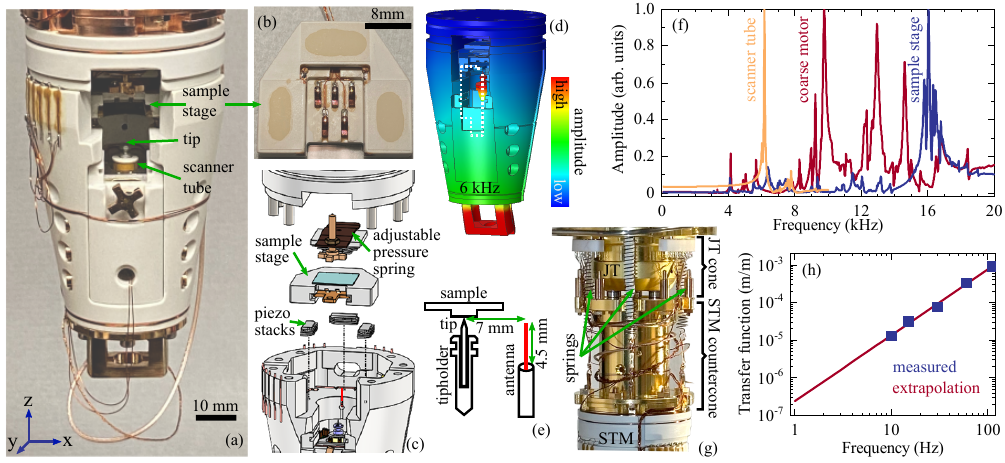}
\caption{\label{The STM} {\bf Scanning tunneling microscope} (a) Photograph of the STM. (b) Photograph of the sample stage with receptacle providing six sample contacts, five by the visible springs and one by contacts in the guiding rails. (c) Exploded drawing of the upper part of the STM with the sample stage, the piezo stacks for $xy$ movement and the adjustable spring for tuning the friction of the piezo stacks on the sample stage. The RF antenna is marked as a short red line. (d) Simulated displacement amplitudes of the STM body at its first eigenmode with frequency 6\,kHz. Dashed white line: mechanical path between tip and sample. (e) Sketch of tip, sample and RF antenna. (f) Measured resonance curves of the scanner tube holding the tip (yellow), the coarse motor for the tip approach (red), and the sample stage (blue). The excitation uses one of the piezo elements adjacent to the component and the detection another one. (g) Photograph of the spring suspension separating the JT stage with its cone and the STM with its countercone by three soft CuBe springs filled with PTFE tape for damping ($f=4$\,Hz, $Q=5$). (h) Transfer function between the top of the STM body and the tunnel junction (blue squares), 300\,K. The excitation uses a mass oscillated by an actuation coil. Its strength is measured by a low frequency acceleration sensor at the top of the STM. The resulting movement between tip and sample is deduced from the response of the STM feedback loop in constant-current mode. The red line is a linear fit within the double-logarithmic plot.
}
\end{figure*}

A schematic is depicted in Fig.~\ref{minimal stiffness setup}a. In the lower part, eight leaf springs (S3) are preloaded by an adjustable horizontal spring S2 providing a tunable force $F_\mathrm{4}$ that opposes the restoring force $F_\mathrm{3}$ from the lower bellow which appears, if the UHV system is displaced vertically from its equilibrium position. Minimizing the total restoring force by tuning S2 leads to a low vertical resonance frequency of the UHV chambers with Al frame. The horizontal restoring forces are compensated by the upper part. Here, the springs S1 pull by the force $F_2$ on the two-dimensional pendulum made of three rods (red bars in Fig.~\ref{fig:minimal stiffness}a)) and the restoring effective spring force of the bellow. The two-dimensional movement of the rods is enabled by three bearings at each joint in a cardanic arrangement (not shown). To avoid friction, we use frictionless bearings (Fig.~\ref{fig:minimal stiffness}b). The air pressure $p$ contributes to the compensating forces by pushing the bellow structure from below via the intentionally larger diameter of this bellow ($F_p$). The springs S1 are tuned until a low resonance frequency of the UHV chambers results in both horizontal directions.
The relatively compact design is shown in Fig.~\ref{fig:minimal stiffness}c.

We optimized the springs by monitoring the resonance frequencies of the UHV chambers including the Al frame via a low frequency acceleration sensor (Wilcoxon Research\textsuperscript{\textregistered} 731-207) mounted to the top of the STM chamber. Figure~\ref{minimal stiffness setup}d shows the resulting oscillations after optimization and initial displacement. Resonance frequencies and quality factors $Q$ are indicated as resulting from a fit of the data. The obtained frequencies of 0.7\,Hz and 1\,Hz for the vertical and horizontal direction, respectively, are below the operation frequency of the PTC (1.1 - 1.6\,Hz) enabling a damping of the excitation.

To analyze the vibrational reduction, we measured the vertical vibrations at the 3\,K PTC stage and the 4\,K cooling plate independently during PTC operation with the acceleration sensor (Fig.~\ref{fig:minimal stiffness}e). These measurements have been performed prior to the final assembly, once in a separate UHV chamber for the PTC stage and once for the cooling plate with minimal stiffness setup installed and optimized, but without the JT stage and the STM. Both measurements used an operation frequency of the PTC  of 1.2\,Hz. The resulting frequency spectra (Fig.~\ref{fig:minimal stiffness}e) reveal a general reduction of vibrational amplitudes by roughly two orders of magnitude. At the PTC frequency, it is still half an order of magnitude with amplitudes of \SI{26}{\micro\meter_{p}} at the PTC stage and \SI{5}{\micro\meter_{p}} at the cooling plate. 
This reduction by a factor of 0.2 is stronger than expected from the transmission curve deduced from the measured resonance frequency and $Q$ factor (Fig.~\ref{fig:minimal stiffness}d) that is only 0.5. We conjecture that the $Q$ factor increases for lower amplitude as confirmed by experiments with amplitudes in the mm range.

 \begin{figure*}[hbt] 
\includegraphics{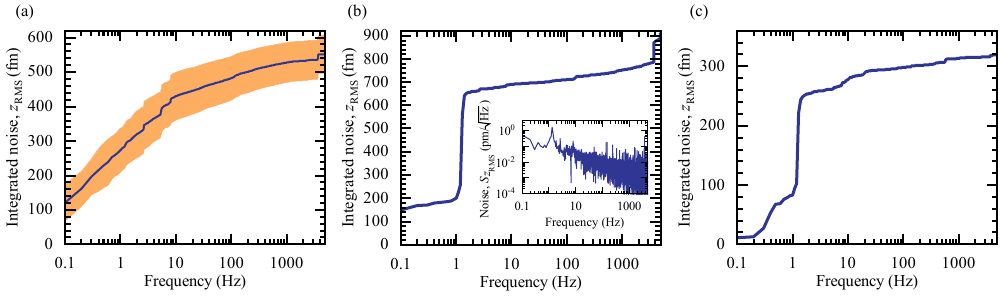}
\caption{\label{znoise} {\bf Vibrations in the tunnel gap ($z$-noise) measured without feedback}. (a) Average of ten curves of the integrated $z$ noise (blue curve) and standard deviation (orange area),  Au(111), W-tip, $V=300$\,mV, $I_\mathrm{stab} = 1$\,nA, $T=1.47$\,K, time interval: 10\,s, sampling rate: 0.1\,ms. (b)  
Single curve of the integrated $z$ noise,  W(100), W-tip, $V=-800$\,mV, $I_\mathrm{stab} = -200$\,pA, $T=1.46$\,K,  time interval: 10\,s, sampling rate: 0.1\,ms, inset: same data in spectral density representation. (c) Integrated $z$ noise, Au(111), W-tip, $V=-480$\,mV, $I_\mathrm{stab} = -1.5$\,nA, $T=1.48$\,K, time interval: 10\,s, sampling rate: 0.1\,ms.
}
\end{figure*}

 \subsection{Vibrational Decoupling of the STM Junction}\label{the microscope}

To further reduce the vibrations towards the tunnel junction of the STM, we centrally employ the transfer function of the STM body. This firstly requires high resonance frequencies, since  the excitation amplitudes strongly decay with frequency (Fig.~\ref{fig:minimal stiffness}e).
Resonance frequencies roughly increase with the ratio of the elastic modulus $E$ to the mass density $\rho$ of the used materials.\cite{Voigtlnder2015} Our microscope is made from the ceramic Shapal\texttrademark\ Hi MSoft, a composition of AlN and BN.\cite{Shapal} The material is light ($\rho=2880$\,kg/m³), stiff ($E=1.8\,\cdot\,10^{11}$\,Pa) and machinable.
Additionally, the STM geometry is optimized for a high resonance frequency using finite element simulations. This leads, e.\,g., to a conical shape of the STM body that reduces the mass at its lower end  (Fig.~\ref{The STM}a). Moreover, it led to a rather compact sample stage made from a 5.5\,mm thick Shapal\texttrademark\ plate (Fig.~\ref{The STM}b) that is clamped to the piezo stacks for $xy$ motion (Fig.~\ref{The STM}c). The remaining lowest resonance frequency of the whole STM in the simulations is the bending mode of the piezo tube used for scanning the tip (5.4\,kHz). The first eigenmode of the STM body has a larger frequency (6\,kHz) with maximum amplitude far away from tip and sample  (Fig.~\ref{The STM}d). The lowest resonance frequency of the sample stage (Fig.~\ref{The STM}b-c) is even larger with 17\,kHz. Importantly, all these frequencies are well above the cutoff of the preamplifier (1.1\,kHz) such that they are not acting on the feedback loop in STM measurements. 

 The simulated frequencies are cross checked experimentally at the various stages in the finalized STM by exciting one piezo element that contacts the corresponding stage and detecting the piezoresponse voltage of another element that contacts this stage  (Fig.~\ref{The STM}f). The observed lowest resonance frequency of the scanner piezo tube (6\,kHz) is likely the calculated bending mode (5.4\,kHz). Also the sample stage shows the first strong resonance at $\sim$\SI{16}{\kilo\Hz}, close to the calculated 17\,kHz. Most importantly, we do not observe any resonance below 4\,kHz, hence avoiding the bandwidth of the preamplifier.

Another important ingredient of the transfer function is the mechanical path between tip and sample that should be minimized to reduce the mutual vibrations. It is marked in Fig.~\ref{The STM}d as a dashed white line amounting to 7\,cm. We measured the mechanical transfer function of the STM using an excitation on the top end of the STM via a mass actuated by an electric coil. The excitation amplitude was measured by a low frequency acceleration sensor (Wilcoxon Research\textsuperscript{\textregistered} 731-207) mounted to the top of the STM. The resulting vibrational amplitude between tip and sample was deduced from the response of the feedback loop in constant-current mode that compensates the low frequency oscillations by more than 98\% as crosschecked by the current signal. The measured amplitude reductions are displayed in Fig.~\ref{The STM}h (blue squares). Extrapolating them down to the PTC operation frequency (1.1 - 1.6\,Hz) reveals an impressive virtual damping by more than six orders of magnitude (red line in Fig.~\ref{The STM}h). 

A final vibrational isolation is provided by three soft CuBe springs installed between the STM and the JT stage (Fig.~\ref{The STM}g). They exhibit a resonance frequency of 4\,Hz and are damped down to $Q=5$ by PTFE tapes within the spirals. They act on higher frequencies such as the higher harmonics of the PTC visible in Fig.~\ref{fig:minimal stiffness}e. For initial cooling of the STM, a Au covered Cu countercone on top of the STM is pressed into a corresponding cone mounted below the JT stage (Fig.~\ref{The STM}g).    
	
Finally, we describe the operational features of the STM. Its sample stage with six sample contacts accepts sample holders with an area of $12\times 12$\,\unit{\milli\square\meter} (Fig.~\ref{The STM}b). They can be laterally moved ($xy$ directions) with the receptacle across a $4\times 4$\,\unit{\milli\square\meter} range. The force on the piezo stacks that move the sample stage is tunable by a CuBe spring with tension that can be adjusted \textit{in-situ} via a screw rotated by a wobble stick. The scanner enables an in-situ  tip exchange as described elsewhere.\cite{Wiebe2004, Muckel2020}   It is glued into a sapphire prism for the coarse approach that is realized by a stick-slip motion of three pairs of piezo stacks pressed onto the long facets of the prism.\cite{Mashoff2009} Each pair of piezo stacks is individually wired to enable a walker type motion. The scan range at base temperature is $1\times 1$\,\unit{\micro\square\meter} for an applied voltage amplitude of \SI{150}{\volt_{p}}.
In addition, a high frequency antenna is placed  7\,mm away from the tip with axis parallel to the tip (Fig.~\ref{The STM}e). It consists of a \SI{50}{\ohm} coaxial cable (Picocoax\texttrademark\, PCX 42 K 10 AK, by Axon Cable), where the outer shield is stripped for the last 4.5\,mm acting as antenna.

\section{Performance}
\label{sec3}
	 
  The performance of the instrument is benchmarked in the following regarding the remaining mechanical and electrical noise in the tunnel junction, energy resolution, thermal drift, and radio frequency response.

\subsection{Vertical Vibrations in the Tunnel Junction}

\begin{figure*} 
\includegraphics{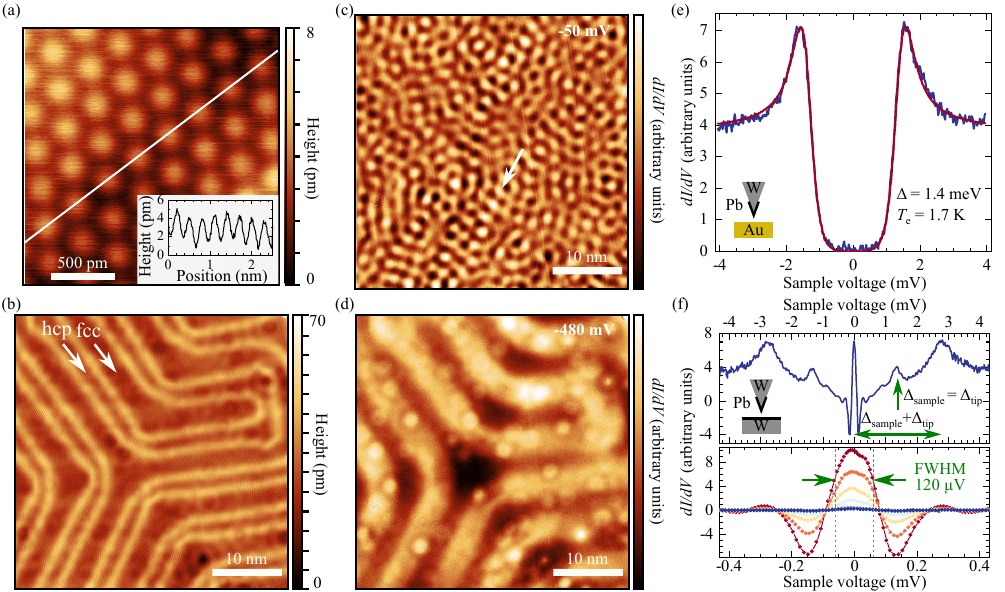}
\caption{\label{STM performance} {\bf STM and STS performance} 
(a) Constant-current image with atomic resolution, Au(111), W-tip, $V=-60$\,mV, $I = -5$\,nA, $T=1.48$\,K, inset: profile along the white line. (b) Constant-current image of Au(111) herringbone reconstruction with marked fcc and hcp areas, W-tip, $V=-50$\,mV, $I = -1$\,nA, $T=1.48$\,K. (c) Differential conductance map $dI/dV(x,y)$ recorded simultaneously with b, $V_\mathrm{mod}=$ \SI{10}{\milli\volt_{p}}. Arrow marks diffractive wave guiding along the herringbone structure.\cite{Libisch2014} (d) $dI/dV(x,y)$,   $V=-480$\,mV, showing confined states in the hcp areas. (e) $dI/dV(V)$ curve (blue) displaying the superconducting gap of a Pb covered W tip (inset),  Au(111), $V_\mathrm{stab}=6$\,mV, $I_\mathrm{stab} = 100$\,pA, $T=1.48$\,K, $V_\mathrm{mod}=$ \SI{20}{\micro\volt_{RMS}}. The red line is a BCS-type fit convolved with the Fermi Dirac distribution revealing the marked parameters.\cite{Liebmann2017,Assig2013} (f) $dI/dV(V)$ curves of Pb covered W tip above Pb covered W(100) (inset), $V_\mathrm{stab}=5$\,mV, $T=1.48$\,K, $V_\mathrm{mod}=$ \SI{15}{\micro\volt_{RMS}},
top: $I_\mathrm{stab} = 200$\,nA with marked peaks (see text), bottom:   $I_\mathrm{stab} = 250$\,nA, $200$\,nA, $150$\,nA, $100$\,nA, $50$\,nA, $10$\,nA with decreasing peak height of the Josephson peak. The FWHM with respect to $dI/dV(V)=0$ (zero line) is marked. }
\end{figure*}

 The stability of the tip-sample distance is the most crucial benchmark for a STM that is cooled by a PTC. It has been measured for a bandwidth of 5\,kHz. 
 
We record the tunnel current $I(t)$ between tip and sample as function of time $t$ without feedback for 10\,s at a sampling rate of 10\,kHz and deduced the vibration in vertical $z$ direction via the $I(z)$ curve that is measured directly before or after recording $I(t)$ (typical average barrier height: 3-5\,eV).\cite{Liebmann2017}
The low pass characteristic of the preamplifier (Femto DLPCA-200) is corrected for according to a calibration measurement that exhibits the 3\,dB point at 1.1\,kHz. 
Via Fourier transformation, we obtain the noise spectrum $S_{z_{\rm RMS}}(f)$ as function of frequency $f$  (Fig.~\ref{znoise}b, inset), which is subsequently integrated revealing the RMS $z$ noise up to frequency $\widetilde{f}$
\begin{equation} \label{eq:1}
    z_{\rm RMS}(\widetilde{f})=\sqrt{\int_{f_0}^{\widetilde{f}} S_{z_{\rm RMS}}(f)^2\, {\rm d}f} 
\end{equation}
with $f_0=0.1$\,Hz.

Displaying the noise as $z_{\rm RMS}(\widetilde{f})$ simplifies to read the total noise in the junction up to a certain frequency and the dominating contributions by the apparent steps. Three curves are displayed in Fig.~\ref{znoise}. They are measured after different cooldowns, on different samples and with different control electronics (RHK R9, Matrix MK1). The curves with 5\,kHz bandwidth all remain below \SI{1}{\pico\meter_{RMS}} of total noise, representing average noise (Fig.~\ref{znoise}a), large noise (Fig.~\ref{znoise}b) and low noise (Fig.~\ref{znoise}c) of the system  that are all achieved after careful tip preparation. In most cases, the oscillation at the operation frequency of the PTC (1.3\,Hz) is still dominating, but favorably reduced to a few 100\,fm. This implies room for further improvement by reducing the minimal stiffness and, hence, the resonance frequencies of the floating UHV system.  Importantly, the total $z$ noise at 5\,kHz bandwidth
is better than for other STMs precooled by PTCs\cite{Meng2019,Geng2023,Huang2022,Ma2023} and also better than for many STMs that are cooled conventionally by externally supplying liquid helium\cite{Liebmann2017,Song2010,Misra2013,Zhang2011,Wiebe2004} (all above \SI{1}{\pico\meter_{RMS}}).
Thus, our STM precooled  by a PTC is competitive to conventional low-temperature STMs.

    \subsection{STM and STS performance}
    \label{subsec3_1}
First STM and STS measurements were performed on a Au(111) surface that has been cleaned by a few cycles of Ar$^+$ sputtering with ion energy 800\,eV and subsequent annealing at  400\,°C for 15\,min. At 1.48\,K, we achieved atomic resolution in constant-current mode with a corrugation of 3\,pm$_{\rm pp}$ and negligible remaining $z$ noise (Fig.~\ref{STM performance}a). 
We also observed the $23\times\sqrt{3}$ herringbone  reconstruction with its alternating hexagonal close-packed (hcp) and face-centered cubic (fcc) stackings\cite{Barth1990}
(Fig.~\ref{STM performance}b), standing electron waves at voltages $V$ well above the guiding potential of the reconstruction (Fig.~\ref{STM performance}c) with some signs of wave guiding by diffraction (arrow),\cite{Libisch2014}
as well as the confined states in the hcp region at $V$ belonging to energies of the surface electrons within the potential troughs of the reconstruction\cite{Chen1998}  (Fig.~\ref{STM performance}d). These data compare favorably to state-of-the-art data for conventionally cooled STMs. We also determined the lateral drift of the tip with respect to the sample by long-term measurements on Au(111) as 18\,pm/h.

\subsection{Energy Resolution}
 \label{subsec3_2}

A central motivation to perform STM at low temperature is the improved energy resolution.\cite{Assig2013,Song2010} For a metallic tip, it is $\Delta E \approx 3.3 k_{\rm B}T_{\rm e}$ due to the broadening by the Fermi-Dirac distribution function.\cite{MORGENSTERN2003} 
Here, $k_{\rm B}$ is the Boltzmann constant and $T_{\rm e}$ is the electronic temperature of the tip that might be different from the phononic temperature $T$, in particular at low $T$. In order to probe $T_{\rm e}$, we use a Au(111) sample and a superconducting tip with its sharp coherence peaks as test objects. Hence, we reverse the role of tip and sample probing the sharpness of the measured coherence peaks. The tip was prepared by indenting a W tip into Pb layers by up to 60\,nm. The resulting $dI/dV(V)$ curve measured by lock-in technique with feedback off is shown in Fig.~\ref{STM performance}e (blue). A BCS-type fit is added (red) deduced from a convolution of the analytic solution of the Maki equation with the Fermi-Dirac distribution function at pair breaking parameter $\xi=0$.\cite{Assig2013,Liebmann2017} The fit parameters are the superconducting gap $\Delta=1.4$\,meV in good agreement with previous STS results\cite{LeDuc1987,Chen2020} and $T_{\rm e}=1.7$\,K, slightly larger than the measured $T=1.48$\,K at the STM. The corresponding energy resolution is $3.3\, k_\mathrm{B} T_\mathrm{e} = 0.48\,\mathrm{meV}$.

The energy resolution can be further improved with a superconducting tip on a superconducting sample.\cite{Rodrigo2004} In that case, it is not limited by $T_{\rm e}$, but only by the voltage noise in the tunnel junction that includes the interactions of the tip with the virtual electromagnetic surrounding depending on the tip shape.\cite{Ast2016}
To probe the remaining voltage noise, we employ a Josephson junction consisting of a Pb covered W tip, vacuum and a Pb covered W(100) crystal. We evaporated Pb at 300\,K onto W(100) that has been prepared by several cycles of annealing at $950^\circ$C in an O$_2$ pressure $p_{O_2}=1\cdot\,10^{-7}$\,mbar and subsequent flash to 2100\,$^{\circ}$C. The W tip was indented into the Pb film to create the superconducting tip. The resulting  $dI/dV(V)$ curve  features prominent peaks at $V$ corresponding to the gap energy of Pb $\Delta_\mathrm{tip}=\Delta_\mathrm{sample}$ and at its double value $\Delta_\mathrm{tip}+\Delta_\mathrm{sample}$ (Fig.~\ref{STM performance}f, top). These peaks are attributed to Andreev reflections and quasiparticle tunneling, respectively.\cite{Ternes2006,Naaman2004} Faint peaks at lower $V$ correspond to multiple Andreev reflections.\cite{Ternes2006} The peak at zero bias is the Josephson peak  caused by Cooper pair tunneling. We crosschecked that it increases superlinear with stabilization current $I_{\rm stab}$ (Fig.~\ref{STM performance}f, bottom).\cite{Liebmann2017} More importantly, the Josephson peak yields a full width at half maximum (FWHM) of \SI{120}{\micro\volt} only, which is significantly smaller than $3.3 k_\mathrm{B} T=0.42\,\mathrm{meV}$ at $T = 1.48\,\mathrm{K}$. The FWHM is, hence, attributed to the remaining voltage noise in the tunnel junction via the cables, partially being thermally induced.  
Indeed, we used the Josephson peak to minimize the voltage noise, e.\,g. 
by low-pass filter boxes, at all electrical lines that enter the STM chamber via feedthroughs except the line for tunneling  current and lines that are grounded during the measurements.
 
	\subsection{High frequency Performance}\label{subsec3_3}

  \begin{figure*} 
\includegraphics{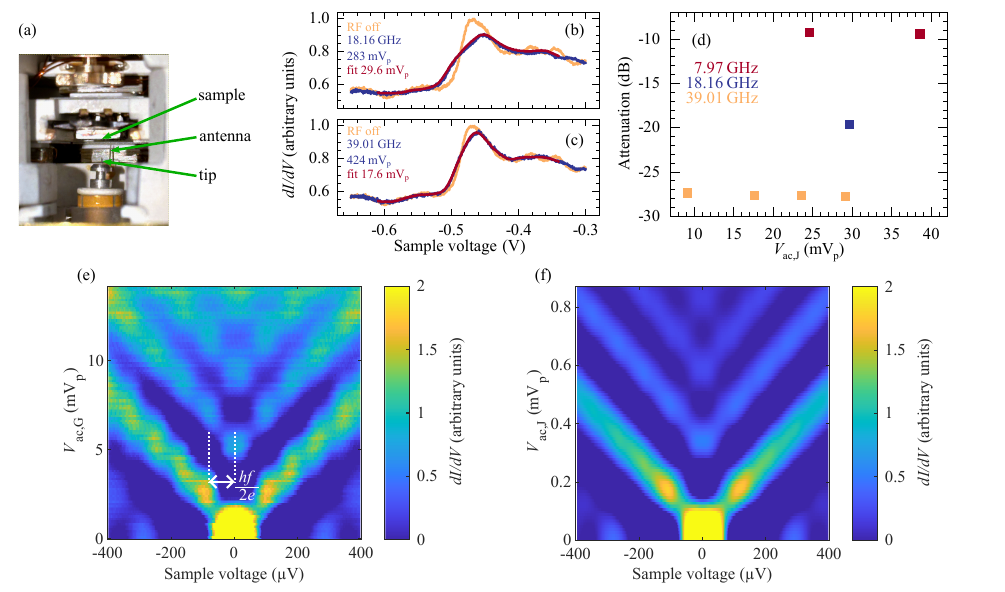}
\caption{\label{RFperformance}   {\bf High frequency performance}. (a) Photograph of microwave antenna, retracted tip, and sample. (b), (c) Differential conductance curves of the Au(111) surface state\cite{Chen1998} recorded without (orange) and with (blue) high frequency as indicated, red line: fit to the blue curve, amplitudes at the generator (blue) and in the tunnel junction resulting from the fit (red) are marked, feedback off, $V_\mathrm{stab}=-650$\,mV, $I_\mathrm{stab} = -1$\,nA, $T=1.50$\,K, $V_\mathrm{mod}=$ \SI{10}{\milli\volt_{p}}. (d) Measured amplitude attenuation from the generator to the tunnel junction for various frequencies and amplitudes in the tunnel junction $V_{\rm ac, J}$.  (e) Differential conductance map of a tip-sample Josephson junction recorded with high frequency voltage, 38.73\,GHz, $V_\mathrm{ac,G}$: amplitude at the high frequency generator, Pb/W(100), Pb covered W tip, $V_\mathrm{stab}=5$\,mV, $I_\mathrm{stab} = 150$\,nA, $T=1.48$\,K, $V_\mathrm{mod}=$ \SI{10}{\micro\volt_{RMS}}. (f) Simulated $dI/dV(V,V_{\rm ac,J})$ at 38.73\,GHz using the Tien-Gordon equation\cite{Tien1963} with the spectrum at $V_{\rm ac,J}=0$\,mV (Fig.~\ref{STM performance}f) as input (eq.~(\ref{eq:TienGordon})). Since the attenuation was not calibrated independently, the $V_{\rm ac,J}$ axis is manually adjusted to e.
}
\end{figure*}

 Implementing high frequency electric fields into a STM tunnel junction enables to probe atomic and molecular dynamics with high spatial control including the superposition and entanglement of spin states.\cite{Baumann2015,Yang2019,Wang2023,Chen2022,Cocker2016} 
 We implemented this option into our PTC based STM via an antenna close to the tip  (Fig.~\ref{RFperformance}a). The antenna is  coupled to a radio frequency (RF) signal generator via multiple \SI{50}{\ohm} flexible coaxial cables.\cite{Seifert2020,Drost2022} 
The generator (Agilent E8254A 1E1 1EA UNJ) is firstly connected to a vacuum feedthrough (Vacom CF16-SMA50-40GHz, Typ K, 2.92\,mm, 50\,Ohm up to 45\,GHz) by a flexible cable (Mini-Circuits CBL-SMSM+). Inside the vacuum, the RF signal is guided to the JT stage by another flexible cable (Elspec Storm Flex 034, length: $\sim 70$\,cm, diameter: 0.93\,mm) that is anchored at 80\,K, 3.6\,K and 1.2\,K. From there to the antenna, we employ a very flexible cable (Picocoax\texttrademark\, PCX 42 K 10 AK, Axon Cable,  length: 22\,cm) connected via a SMPM connector (Cinch Connectivity Solutions 125-2592-001, Amphenol 925-124P-51S). For the last 4.5\,mm, the shielding is removed to realize the antenna (Fig.~\ref{The STM}e).

 To determine the RF amplitude in the tunnel junction $V_\mathrm{ac,J}$, we use the onset of the Au(111) surface state at $V=-480$\,mV (Fig.~\ref{RFperformance}b-c, orange curve).\cite{Chen1998} It is broadened by the oscillating tip-sample voltage originating from the RF signal (Fig.~\ref{RFperformance}b-c, blue curve).\cite{Paul2016} Fitting the resulting $dI/dV(V)$ by convolving the $dI/dV(V)$ curve without RF and a normalized arcsine distribution (Fig.~\ref{RFperformance}b-c, red curve) returns  $V_\mathrm{ac,J}$ as the only fit parameter.\cite{Paul2016, Seifert2020} The successful fit enables to deduce the attenuation or transfer function 
 \begin{eqnarray}
 T=\mathrm{20\ log_{10}}(V_\mathrm{ac,J}/V_\mathrm{ac,G})
 \label{eq:one}
\end{eqnarray}
 with the amplitude $V_\mathrm{ac,G}$  provided by the generator (Fig.~\ref{RFperformance}d). 
 As usual, we find that $T$ is independent of amplitude and decays with frequency
exhibiting -28\,dB at 40\,GHz, the highest frequency of the generator. This attenuation is comparable to other STMs with conventional cooling.\cite{Hwang2022,Seifert2020}

As a recently developed application of the RF voltage, we demonstrate microwave assisted Cooper pair tunneling.\cite{Peters2020,Kot2020}
 We use a Josephson junction of a Pb covered W tip, vacuum, and Pb covered W(100) as probed without RF excitation in Fig.~\ref{STM performance}f. The $dI/dV(V)$  with RF excitation at frequency $f\approx 40$\,GHz is shown as function of $V_{\rm ac, G}$ in Fig.~\ref{RFperformance}e. The Josephson peak develops into multiple peaks in $V$-direction with distance $\Delta V ={hf}/{2e}=80\,\mu$V ($e$: elementary charge) due to multiphoton excitations of the Cooper pairs across the junction (double arrow in Fig.~\ref{RFperformance}e). The peak sequence exhibits a V-shape in the ($V$, $V_{\rm ac,G}$) plane reflecting the relation between amplitude and the number of photons.\cite{Peters2020} 
Quantitatively, this has been described successfully by the Tien-Gordon equation that also includes other subgap features \cite{Tien1963, Falci1991, Roychowdhury2015}

 \begin{eqnarray}
 I_{f}(V,\,V_\mathrm{ac,J})=
 \sum^{\infty}_{n=-\infty}
 J_\mathrm{n}^2\left(\frac{2eV_\mathrm{ac,J}}{hf}\right)I_\mathrm{0}\left(V-\frac{nhf}{2e}\right)
\label{eq:TienGordon}
\end{eqnarray}
 
with the $n^{\rm th}$ Bessel function  $J_\mathrm{n}$, the Planck constant $h$, and the dc tunnelling current without RF excitation $I_\mathrm{0}$. The result is displayed in Fig.~\ref{RFperformance}f matching the experimental data nicely.
The attenuation can be determined from the adaption of the $V_{\rm ac,J}$ axis to the experiment being $T=-25$\,dB. This is in good agreement with the data of Fig.~\ref{RFperformance}d regarding the fact that $T$ typically exhibits some temporal fluctuations.\cite{Paul2016}

\section{Conclusion}\label{conclusion}
We have presented a scanning tunneling microscope that operates at 1.5\,K without supplying external liquids. It employs a combination of a two-stage pulse tube cooler (PTC) and a Joule-Thomson cooling stage and robustly achieves a stability in $z$ direction below 1\,pm$_{\rm RMS}$ at 5\,kHz bandwidth, partly going down to 300\,fm$_{\rm RMS}$. The efficient vibrational decoupling consists of bellows that are tuned below the resonance frequency of the PTC by counteracting spring forces, a decoupling of the PTC stages from cooling plates for the STM area by thin Cu wires, a spring suspension of the STM, and a STM design providing a very small vibrational transfer at low frequency. We believe that this is a significant breakthrough since the decoupling stages can be very similarly applied for other secondary cooling cycles as, e.\,g.,   
dilution refrigerators.\cite{Song2010,Assig2013,vanWeerdenburg2021}
The system design also directly enables the implementation of a superconducting magnet around the STM.

\begin{acknowledgments}
We are grateful to J. Falter, C. Ast, S. Hild, M. Liebmann and F. Muckel for helpful discussions and J. Beeker as well as J.-M. Grigoleit for technical and instrumental support and the German Ministry for Economic Affairs and Climate Action (BMWK) via the  ZIM project ZF 4367101SY6 for financial support.
\end{acknowledgments}


%

\end{document}